\documentclass[fleqn,10pt]{wlscirep}
\usepackage[utf8]{inputenc}
\usepackage[T1]{fontenc}
\title
{
Emerging symmetric strain response and 
weakening nematic fluctuations in strongly hole-doped iron-based superconductors
}

\author[1]{P.~Wiecki}
\author[1]{M.~Frachet}
\author[1]{A.-A.~Haghighirad}
\author[1]{T.~Wolf}
\author[1]{C.~Meingast}
\author[1]{R.~Heid}
\author[1,2,*]{A.~E.~B\"{o}hmer}
\affil[1]{Karlsruhe Institute of Technology, Institute for Quantum Materials and Technologies, 76021 Karlsruhe, Germany}
\affil[2]{Institut f\"ur Experimentalphysik IV, Ruhr-Universit\"at Bochum, 44801 Bochum, Germany}

\affil[*]{anna.boehmer@kit.edu}

\begin{abstract}
Electronic nematicity is often found in unconventional superconductors, suggesting its relevance for electronic pairing. 
In the strongly hole-doped iron-based superconductors, the symmetry channel and strength of the nematic fluctuations, as well as the possible presence of long-range nematic order, remain controversial.
Here, we address these questions
using transport measurements under elastic strain.
By decomposing the strain response into the appropriate symmetry channels, 
we demonstrate the emergence of a giant in-plane symmetric
contribution, associated with the growth of both strong electronic correlations and the sensitivity
of these correlations to strain. 
We find weakened remnants of the nematic fluctuations that are present at optimal doping, but no change in the 
symmetry channel of nematic fluctuations with hole doping. Furthermore, we find no evidence
for a nematic-ordered state in the AFe$_2$As$_2$ (A = K, Rb, Cs) superconductors. 
These results revise the current understanding of nematicity in hole-doped iron-based superconductors.

\end{abstract}
\begin{document}

\flushbottom
\maketitle

\thispagestyle{empty}

\section*{Introduction}

Nematicity, the breaking of rotational symmetry by electronic interactions, has by now been observed in a variety of unconventional superconductors. 
In addition to iron-based superconductors with almost ubiquitous nematicity\cite{Kuo958,Fernandes_2014}, nematicity is discussed in the context of cuprate high-$T_c$ superconductors\cite{Kivelson_1998,Hinkov_2008,Daou_2010,Murayama_2019,Sato_2017,Achkar_2016,PhysRevB.92.224502,Lawler_2010,Vojta_2009,Auvray_2019}, heavy-fermion superconductors\cite{Ronning_2017,Helm_2020}, intercalated Bi$_2$Se$_3$ topological superconductors\cite{Yonezawa_2016,Shen_2017,Hecker_2018,Cho_2020} and even twisted bilayer graphene\cite{cao2020nematicity}. 
Furthermore, it has been theoretically suggested that nematic fluctuations may enhance pairing and therefore be an important ingredient for high-$T_c$ superconductivity \cite{Eckberg_2019,Malinowski_2020,Lederer_2015,Lederer_2017}. 

In iron-based superconductors,
nematicity has been intensively studied in the vicinity of the parent compound BaFe$_2$As$_2$ 
 because the stripe-type antiferromagnetic ground state inherently breaks the $C_4$ rotational symmetry of the high-temperature tetragonal phase \cite{NandiPRL2010,Chu_2012,Kuo958,Fernandes_2014}, corresponding to a nematic degree of freedom. The accompanying structural distortion is in the $B_{2g}$ channel of the tetragonal $D_{4h}$ point group.
In electron-doped BaFe$_2$As$_2$, the structural distortion occurs at a higher temperature than the antiferromagnetic state, creating a long-range, nematic-ordered phase.
Nematic fluctuations of $B_{2g}$ symmetry are observed near optimal doping in both hole- and electron-doped BaFe$_2$As$_2$\cite{Gallais_2013,B_hmer_2014}. 
Such fluctuations have frequently been studied using elastoresistance, the strain dependence of electrical resistivity\cite{Chu2012,Kuo958,Ikeda_2021}. 

The parent compound BaFe$_2$As$_2$ nominally has a $3d^6$ Fe configuration. 
With hole doping the Fe electron configuration begins to approach the half-filled $3d^5$, where a Mott insulating state is expected theoretically\cite{de_Medici_2009,Misawa_2012,de_Medici_2014}. 
Indeed, signatures of strong electronic correlations and orbital-selective Mott behavior have been observed in the isoelectronic $3d^{5.5}$ series AFe$_2$As$_2$ (A = K, Rb, Cs), including an enhanced Sommerfeld coefficient and signs of a coherence-incoherence crossover\cite{Hardy_2013,Hardy_2016,Wu_2016}.
On the basis of the strong increase of electronic correlations with increasing alkali ion size in AFe$_2$As$_2$ (A = K, Rb, Cs),
it has further been proposed that these compounds lie near a QCP associated with the suppression of a (thus far unobserved) ordered phase, possibly related to the $3d^5$ Mott insulator\cite{Eilers_2016,Zhang_2018}. 
Furthermore, the electronic correlations in AFe$_2$As$_2$ have been found to be highly sensitive to in-plane strain\cite{Eilers_2016,Hardy_2016}.

\begin{figure*}[t]
\centering
\includegraphics[width=\textwidth]{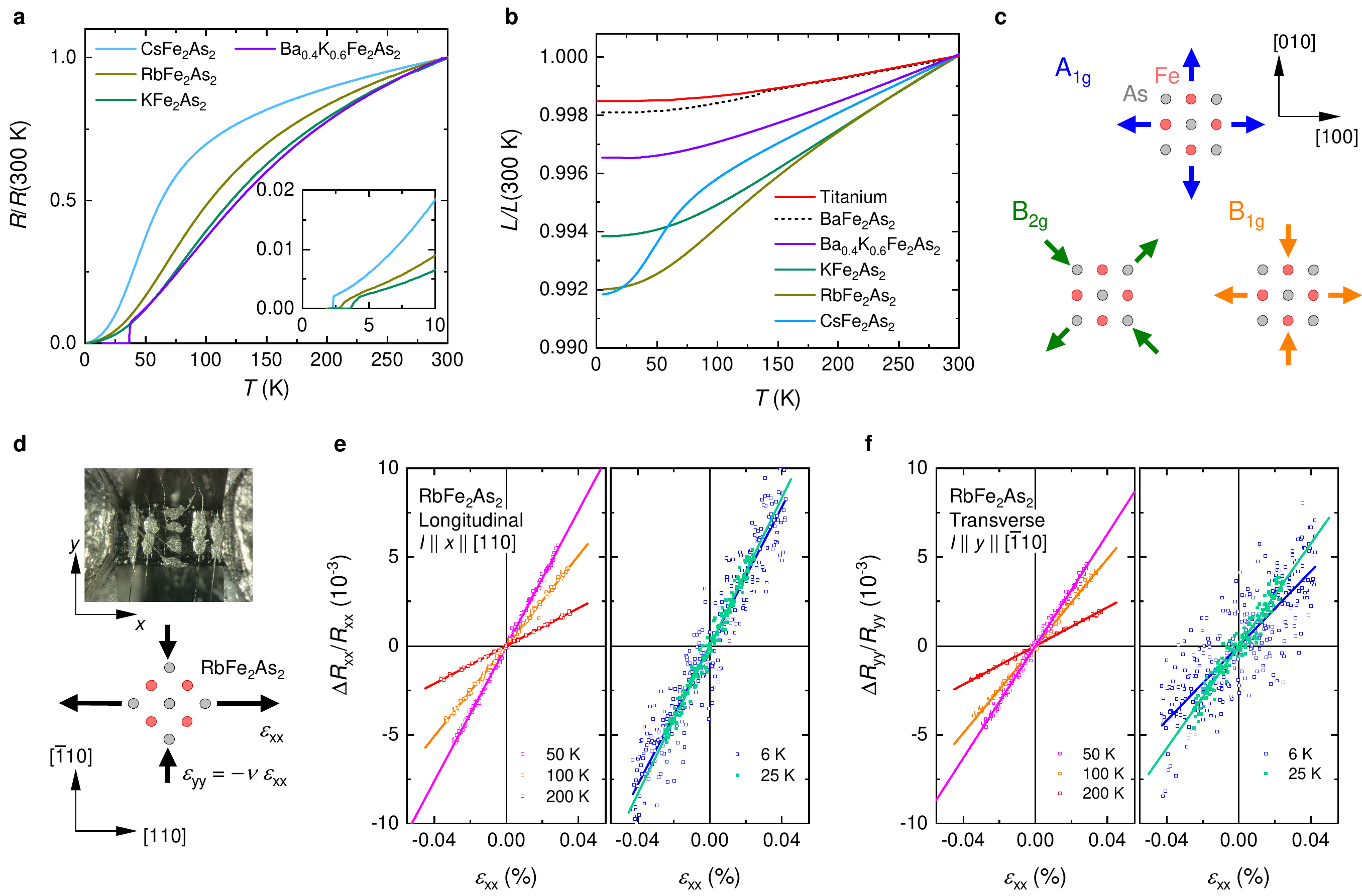}
\caption{\textbf{Basic material properties and experimental setup.} \textbf{a} Normalized electrical resistance as a function
of temperature. Inset: zoom of superconducting transitions. \textbf{b} In-plane sample length changes as a function of 
temperature\cite{Hardy_2016}.
Titanium and BaFe$_2$As$_2$\cite{Meingast_2012} are shown for comparison. \textbf{c} Sketch of the deformations in the Fe plane corresponding to the indicated irreducible representations.
\textbf{d} Photograph of a RbFe$_2$As$_2$ sample mounted between titanium plates in the strain cell. 
The uniaxial stress axis is defined to be $x$. In this example, strain is
applied along [110]
and the sample experiences $A_{1g}+B_{2g}$ strain.
$\nu$ is the in-plane Poisson ratio of the sample. 
\textbf{e}-\textbf{f} Longitudinal and transverse elastoresistance curves at indicated temperatures. Lines 
are linear fits. }
\label{fig:intro}
\end{figure*}

The fate  of nematicity in the strongly hole-doped iron-based superconductors remains controversial. 
In the Ba$_{1-x}$K$_x$Fe$_2$As$_2$ series, the elastic softening associated with the $B_{2g}$ nematic fluctuations decreases with hole doping and is no longer observed for  $x\geq 0.82$ \cite{B_hmer_2014}. 
Several recent studies have suggested a change to nematic fluctuations of $B_{1g}$ symmetry in the $3d^{5.5}$ compounds, in contrast to the pervasive $B_{2g}$ nematic fluctuations observed at optimal doping\cite{li2016reemergeing,Ishida6424,Liu_2019,Onari_2019,Wang_2019,Borisov_2019,Moroni_2019}. 
Furthermore, an ordered $B_{1g}$ nematic phase has been proposed in RbFe$_2$As$_2$ based on a maximum in the elastoresistance\cite{Ishida6424} and an asymmetry observed in low-temperature STM\cite{Liu_2019}.
However, the study of nematic fluctuations in these compounds by elastoresistance is complicated by the emergence of the strong electronic correlations, which create a significant contribution in the in-plane symmetric $A_{1g}$  channel\cite{PhysRevLett.125.187001}.
In such a case, it is essential to properly decompose the contributions
in different symmetry channels by comparing the resistance changes both longitudinal and transverse to the stress axis\cite{PhysRevB.88.085113,PhysRevB.96.205133,PhysRevLett.125.187001}. 
In addition, the extreme thermal expansion of these samples (Fig. \ref{fig:intro}\textbf{b}) poses an experimental challenge for controlled measurements under elastic strain\cite{PhysRevLett.125.187001}. 

In this work, we present comprehensive elastoresistance measurements on the hole-doped iron-based superconductors Ba$_{0.4}$K$_{0.6}$Fe$_2$As$_2$ and AFe$_2$As$_2$ (A = K, Rb, Cs)
carried out in a piezoelectric-based strain cell capable of full control over the strain state of the sample. 
We find a monotonic increase of the $A_{1g}$ response with hole doping and increasing alkali ion size.
In contrast, the $B_{2g}$ elastoresistance, a measure of the $B_{2g}$ nematic susceptibility,
weakens with hole-doping. The $B_{1g}$ elastoresistance shows no sign of a divergence, thus there is no indication of the emergence of a $B_{1g}$-type nematic instability with increasing hole doping.
Furthermore, we find a clear correspondence between the low-temperature elastoresistance and the electronic Gr\"{u}neisen parameter if the elastoresistance coefficient is defined based only on the temperature dependent part of the resistance. 

\begin{figure*}[t]
\centering
\includegraphics[width=\textwidth]{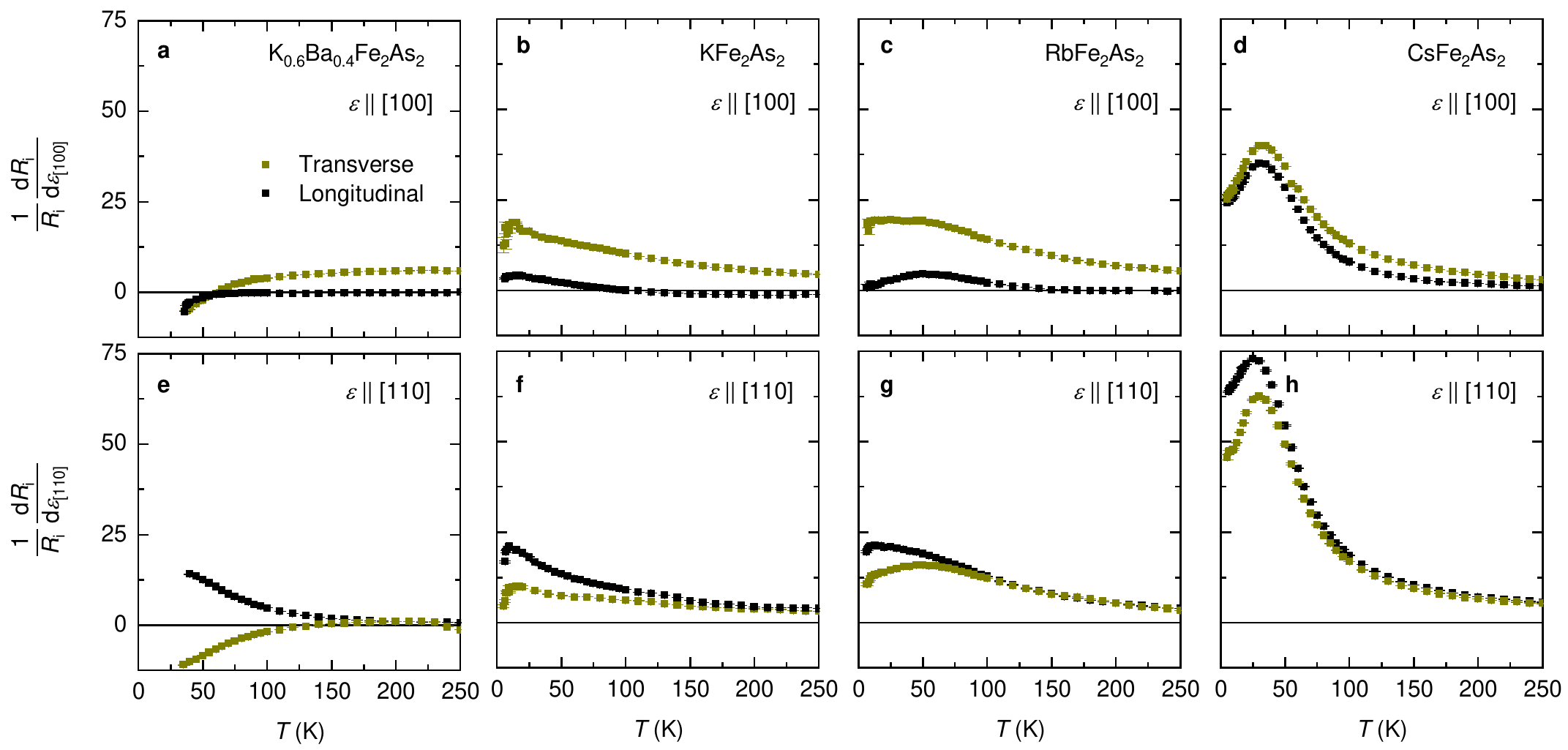}
\caption{\textbf{Elastoresistance of strongly hole-doped iron based superconductors.} 
Longitudinal and transverse elastoresistance with the strain direction $\epsilon_{xx}\|[100]$ (\textbf{a}-\textbf{d}) and 
$\epsilon_{xx}\|[110]$ (\textbf{e}-\textbf{h}) for indicated samples arranged by column. Error bars indicate one standard deviation.
}
\label{fig:rawdata}
\end{figure*}

\section*{Results}

\subsection*{Elastoresistance measurements}

The freestanding electrical resistance of our samples is displayed in Fig. \ref{fig:intro}\textbf{a}.
For elastoresistance measurements, the sample is mounted in a strain cell composed of titanium (Fig. \ref{fig:intro}\textbf{d}). 
Fig. \ref{fig:intro}\textbf{b} displays the length of the samples as a function of temperature, along with titanium for comparison. 
We see that the hole-doped samples shrink much more than the titanium apparatus on cooling. 
Thus, the samples can be maintained near neutral strain on cooling by adjusting the distance between the sample mounting plates in the strain cell based on the thermal expansion difference between the sample and titanium (see Methods).
We place eight electrical contacts on the sample so that both longitudinal and transverse elastoresistance can be measured in the same experimental run (Fig. \ref{fig:intro}\textbf{d}). 
Raw elastoresistance data on RbFe$_2$As$_2$ obtained in this way are shown in Figs. \ref{fig:intro}\textbf{e},\textbf{f} as an example.

\begin{figure}[t]
\centering
\includegraphics[width=0.5\columnwidth]{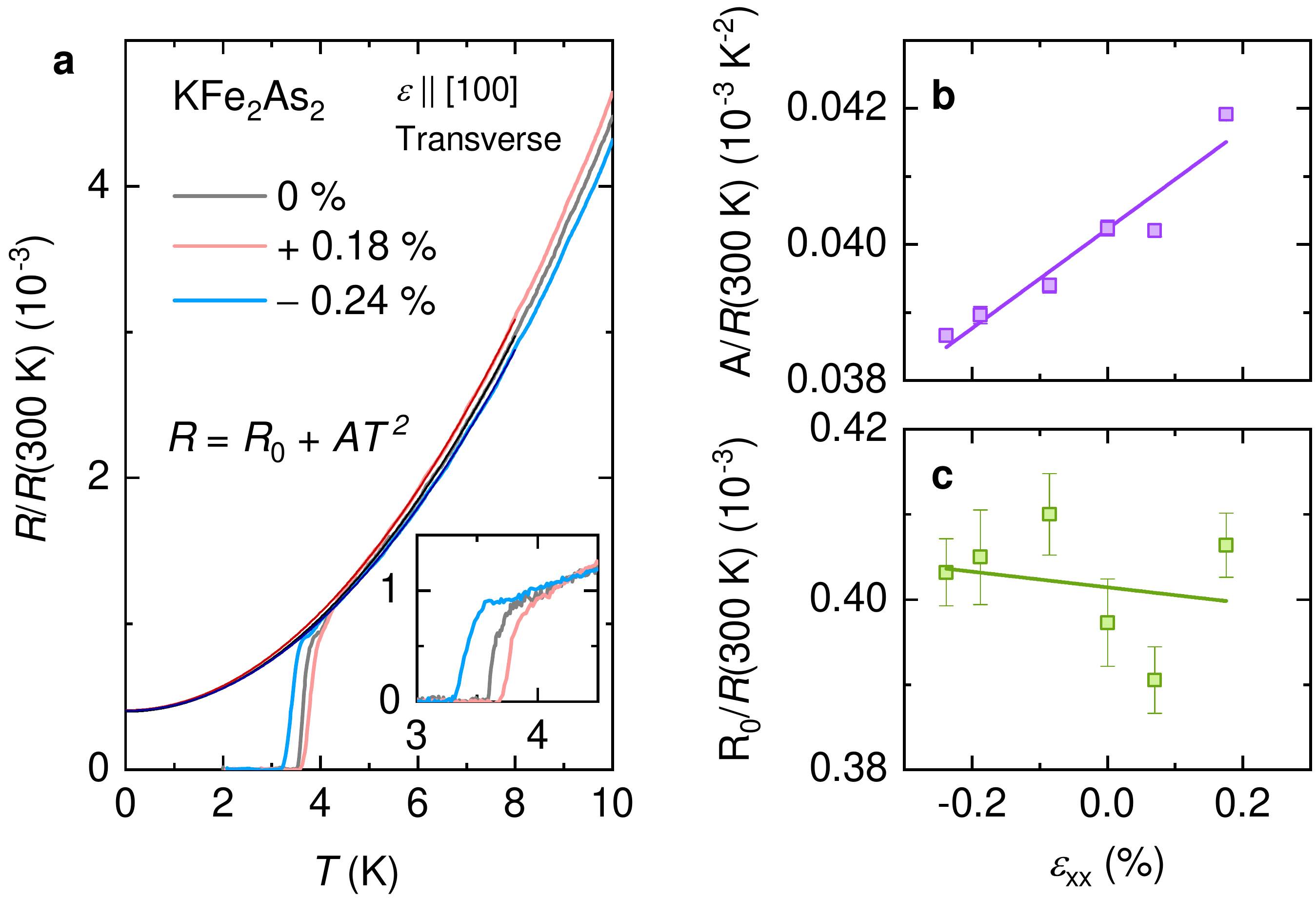}
\caption{
\textbf{Resistance at fixed strain.} 
\textbf{a}, Normalized resistance of KFe$_2$As$_2$ as a function of temperature at fixed strains, along with fits to the indicated fitting function. 
Inset: zoom of superconducting transitions. 
\textbf{b}, Strain dependence of the fit parameter $A$, measuring electronic correlations and effective mass. 
\textbf{c}, Strain dependence of the fit parameter $R_0$, the residual resistance. 
Lines in \textbf{b}, \textbf{c} are linear fits. 
}
\label{fig:AvsStrain}
\end{figure}
\begin{figure}%[t]
\centering
\includegraphics[width=0.5\columnwidth]{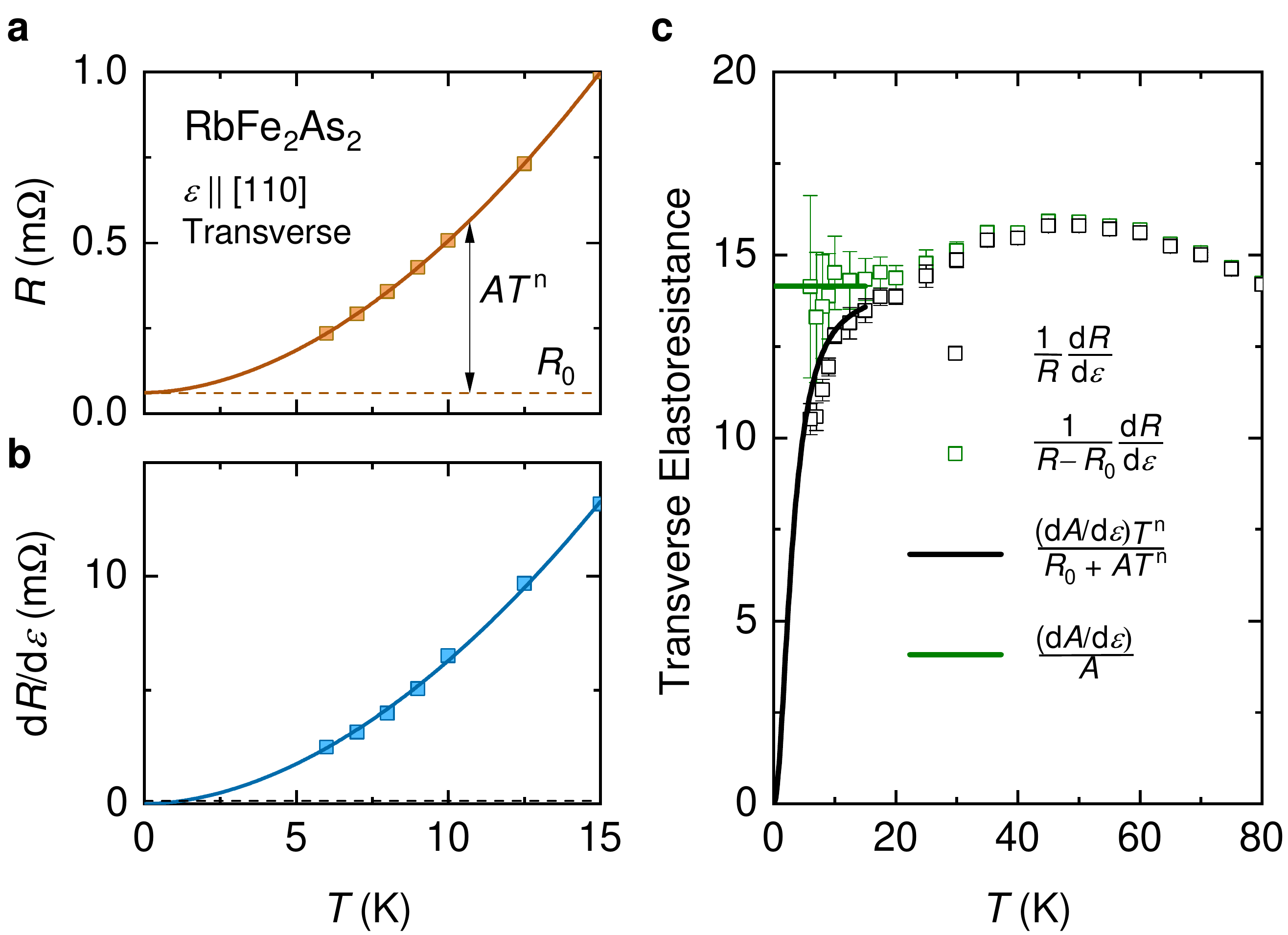}
\caption{\textbf{Redefinition of of elastoresistance based on the temperature-dependent resistance contribution
}
\textbf{a}, The resistance of a particular RbFe$_2$As$_2$ crystal 
at temperatures where elastoresistance was measured. The solid line is a 
fit of these points to $R=R_0+AT^n$ (here $n=1.84$). The dashed line represents the residual resistance $R_0$. 
\textbf{b}, The strain derivative of the sample resistance at selected temperatures. 
The solid line is a fit of these points to $dR/d\epsilon=B+CT^n$ (again $n=1.84$). $B$ is consistent with zero
within uncertainty. The dashed line gives an upper limit on $B$, from the fit uncertainty. 
\textbf{c}, Elastoresistance calculated as $m=(1/R)dR/d\epsilon$ (black) and $\bar{m}=[1/(R-R_0)]dR/d\epsilon$ (green).
The error bars on $[1/(R-R_0)]dR/d\epsilon$ reflect the uncertainty in the estimation of $R_0$.
The solid lines show the same quantities instead calculated from the fitted functions in \textbf{a},\textbf{b}.
}
\label{fig:LTC}
\end{figure}

In Figs. \ref{fig:rawdata} \textbf{a}-\textbf{h} we present the longitudinal and transverse elastoresistance 
 for all samples. Here, the elastoresistance is defined, for the moment, simply as
$m_{ii,xx}=(1/R_{ii})dR_{ii}/d\epsilon_{xx}$, where $i=x$ corresponds to the longitudinal and $i=y$ corresponds to the transverse elastoresistance. 
The strain is applied along $x\|$[100] (Figs. \ref{fig:rawdata} \textbf{a}-\textbf{d}) 
or $x\|$[110] (Figs. \ref{fig:rawdata} \textbf{e}-\textbf{h}). When strain is applied along [100] the transverse elastoresistance is larger than the longitudinal over most of the temperature range for all samples. In contrast, for strain applied along [110]
we observe that the longitudinal and transverse elastoresistance are equal at high temperature, while the longitudinal becomes larger at low temperature, for all samples. 
In the case of Ba$_{0.4}$K$_{0.6}$Fe$_2$As$_2$ with strain along [110] 
(Fig. \ref{fig:rawdata} \textbf{e}), the longitudinal and transverse elastoresistance have opposite sign. This is the expected
behavior for dominant $B_{2g}$ nematic fluctuations\cite{PhysRevB.88.085113}. In contrast, in the $3d^{5.5}$ superconductors AFe$_2$As$_2$ (A=K, Rb, Cs) both the longitudinal and transverse elastoresistance are positive at all temperatures. 

We note significant drops in the elastoresistance on decreasing temperature below $\sim15$ K for both transverse and longitudinal, most notably in KFe$_2$As$_2$ and RbFe$_2$As$_2$
(Figs. \ref{fig:rawdata} \textbf{b},\textbf{c},\textbf{f},\textbf{g}). These anomalies are rather broad (as also visible in Fig. \ref{fig:LTC}\textbf{c} with an enhanced scale) and we do not associate them with a phase transition, since no hint of a phase transition is seen in either $R(T)$ or the thermal expansion $L(T)$ (Fig. \ref{fig:intro}), nor indeed in the respective temperature derivatives $dR(T)/dT$ and $dL(T)/dT$ (Supplemental Figure). Rather, as explained below in relation to Fig. \ref{fig:LTC}, the drop in elastoresistance results from the sample's
resistance becoming comparable to its residual resistance. Note, however, that the broad peak in CsFe$_2$As$_2$ at $\sim 35$ K is a pronounced manifestation of the coherence-incoherence crossover in this material.

\subsection*{Physical mechanism for strain dependence of resistance}
To obtain a better physical understanding of the elastoresistance, it is useful to take a complementary view and study the 
resistance as a function of temperature at fixed strain. 
This is shown in Fig. \ref{fig:AvsStrain}\textbf{a}, taking  KFe$_2$As$_2$ as an example.
At low temperatures, the data can be fit to the standard Fermi liquid form $R=R_0+AT^2$.  
The coefficient $A$ is a measure of electronic correlations and effective mass, as given by the Kadowaki-Woods relation for a Fermi liquid $A\propto\gamma^2$, where $\gamma$ is the Sommerfeld coefficient.
The fit parameters $R_0$ and $A$ are shown as a function of strain in Figs. \ref{fig:AvsStrain}\textbf{b},\textbf{c}.
We observe that the coefficient $A$ increases linearly with strain, while the 
residual resistance $R_0$ is strain independent within experimental uncertainty.
We have similarly confirmed that $R_0$ is also strain independent in CsFe$_2$As$_2$.

From the Kadowaki-Woods relation, the increase of $A$ with strain is  consistent with the large positive $d\gamma/d\epsilon$
inferred from thermal expansion measurements\cite{Eilers_2016,Hardy_2016} (see Methods) and LDA+DMFT calculations\cite{PhysRevB.92.195128}.
The increase of $A$ under tensile strain ($\epsilon>0$) is also consistent with the increase of the Sommerfeld coefficient in unstrained AFe$_2$As$_2$ (A = K, Rb, Cs)
(Fig. \ref{fig:intro}\textbf{a}) from A = K to A = Cs. Namely, the increasing alkali ion size from K to Cs creates a negative chemical pressure, 
reducing bandwidths and enhancing correlations.
Consistently, hydrostatic pressure reduces $A$ in KFe$_2$As$_2$\cite{Taufour_2014}.
Furthermore, we observe that $T_c$ increases with tensile strain. The strain dependence of $T_c$ is in rough quantitative agreement
with inferences based on thermodynamic measurements\cite{Eilers_2016} (see Methods).

In RbFe$_2$As$_2$ and CsFe$_2$As$_2$, we find that the resistance at low temperature can be fit to $R=R_0+AT^n$,
but with an exponent $n<2$ (Fig. \ref{fig:LTC}\textbf{a}). 
This is consistent with the non-Fermi liquid behavior seen at low temperature in these samples by NMR\cite{Zhang_2018}.
In Fig. \ref{fig:LTC}\textbf{b}, we show the derivative $dR/d\epsilon$ as a function of temperature for the case of RbFe$_2$As$_2$ with strain $\|[110]$ as a representative example. Taking $R_0$ to be independent of strain, $dR/d\epsilon=(dA/d\epsilon)T^n$. Consistently, we find that $dR/d\epsilon$ can be fit to $B+CT^n$ with the same exponent $n$ as the resistance and an intercept $B$
consistent with zero. 
In this situation, if the elastoresistance is simply defined as $m=(1/R)dR/d\epsilon$ as is commonly done (see Fig. \ref{fig:rawdata}), we obtain $m=(dA/d\epsilon)T^n/(R_0+AT^n$). Clearly, $m\to0$ as $T\to0$, as shown in Fig. \ref{fig:LTC}\textbf{c}. As a consequence, the elastoresistance starts to drop on cooling when the residual resistance becomes a significant fraction of the total resistance below 15 K, as noted earlier (Fig. \ref{fig:rawdata}).
If instead the elastoresistance is redefined based on only the temperature dependent contribution to the resistance as $\bar{m}=[1/(R-R_0)]dR/d\epsilon$, we obtain $\bar{m}\to(dA/d\epsilon)/A$ at low temperature (Fig. \ref{fig:LTC}\textbf{c}). 
Note that these arguments are valid for any $n$.
A similar approach has been applied in the analysis of elastoresistance data in YbRu$_2$Ge$_2$\cite{Rosenberg7232}.

The redefined elastoresistance $\bar{m}$ produces more physically meaningful results at low temperature. For example, in the special case of $n=2$ where $A\propto\gamma^2$, $\bar{m}(T\to0)$ is given by
$(dA/d\epsilon)/A=(d\gamma^2/d\epsilon)/\gamma^2=2(d\gamma/d\epsilon)/\gamma$, which 
is proportional to the Gr\"{u}neisen parameter $\Gamma\equiv\-(d\gamma/d\epsilon)/\gamma$ for the Sommerfeld coefficient $\gamma$.  $\Gamma$ is a measure of the strain dependence of an energy scale. Its divergence is a central characteristic of a strain-tuned quantum critical point\cite{Kuechler2004-PRL93.096402,Garst2005-PhysRevB.72.205129,Meingast_2012}, as previously discussed in these materials \cite{Eilers_2016}.

Having understood the longitudinal and transverse elastoresistance data, we now calculate the symmetry-decomposed elastoresistance coefficients $\bar{m}_\alpha$,  where $\alpha$ represents the irreducible representation of the tetragonal $D_{4h}$ point group. These coefficients determine the resistance changes associated with the pure symmetric strains illustrated schematically in Fig. \ref{fig:intro}\textbf{c}.
In terms of the experimental data, $\bar{m}_{A_{1g}}$ is proportional to the sum of the longitudinal and transverse elastoresistance, while $\bar{m}_{B_{1g}}$ ($\bar{m}_{B_{2g}}$) is proportional to the difference when $\epsilon\|[100]$ ($\epsilon\|[110]$) (see Methods). 
The resulting symmetry-decomposed elastoresistance coefficients are shown in Fig. \ref{fig:irrep}.
The in-plane symmetric coefficient $\bar{m}_{A_{1g}}$ (Fig. \ref{fig:irrep}\textbf{a}) strongly increases in magnitude  with hole doping from Ba$_{0.4}$K$_{0.6}$Fe$_2$As$_2$ to KFe$_2$As$_2$ and further with tensile chemical pressure from KFe$_2$As$_2$ to CsFe$_2$As$_2$. Furthermore, $\bar{m}_{A_{1g}}$ of CsFe$_2$As$_2$ has a strong 
temperature dependence, with an increase on cooling and a broad peak near 30 K. 
This temperature dependence is reminiscent of the coherence-incoherence crossover
observed in the thermal expansion coefficient of this material\cite{PhysRevLett.125.187001,Hardy_2016}.
The coefficients $\bar{m}_{B_{2g}}$ (Fig. \ref{fig:irrep}\textbf{b})
and $\bar{m}_{B_{1g}}$ (Fig. \ref{fig:irrep}\textbf{c}) are significantly smaller in magnitude than $\bar{m}_{A_{1g}}$, except for the case of $\bar{m}_{B_{2g}}$ in Ba$_{0.4}$K$_{0.6}$Fe$_2$As$_2$. For all samples, $\bar{m}_{B_{2g}}$ is near zero at high temperature and displays a divergent increase on cooling. 
The positive sign of $\bar{m}_{B_{2g}}$ is opposite to that of BaFe$_2$As$_2$, 
consistent with observed sign-change of resistivity anisotropy upon K substitution in Ba$_{1-x}$K$_{x}$Fe$_2$As$_2$\cite{Blomberg_2013}. 
We also note that both $\bar{m}_{B_{1g}}$ and $\bar{m}_{B_{2g}}$ of Ba$_{0.6}$K$_{0.4}$Fe$_2$As$_2$ are similar in magnitude and 
sign to CaKFe$_4$As$_4$, which is isoelectronic to Ba$_{0.5}$K$_{0.5}$Fe$_2$As$_2$\cite{boehmer2020evolution,Terashima_2020}. 
The coefficient $\bar{m}_{B_{1g}}$ is non-zero
at high temperature and decreases roughly linearly on cooling in RbFe$_2$As$_2$
and KFe$_2$As$_2$. In CsFe$_2$As$_2$ and Ba$_{0.4}$K$_{0.6}$Fe$_2$As$_2$ we observe an upturn at low temperature. 
The temperature dependence of $\bar{m}_{B_{1g}}$ shows no sign of a divergence in any of our samples.

\section*{Discussion}

\begin{figure*}[t]
\centering
\includegraphics[width=\textwidth]{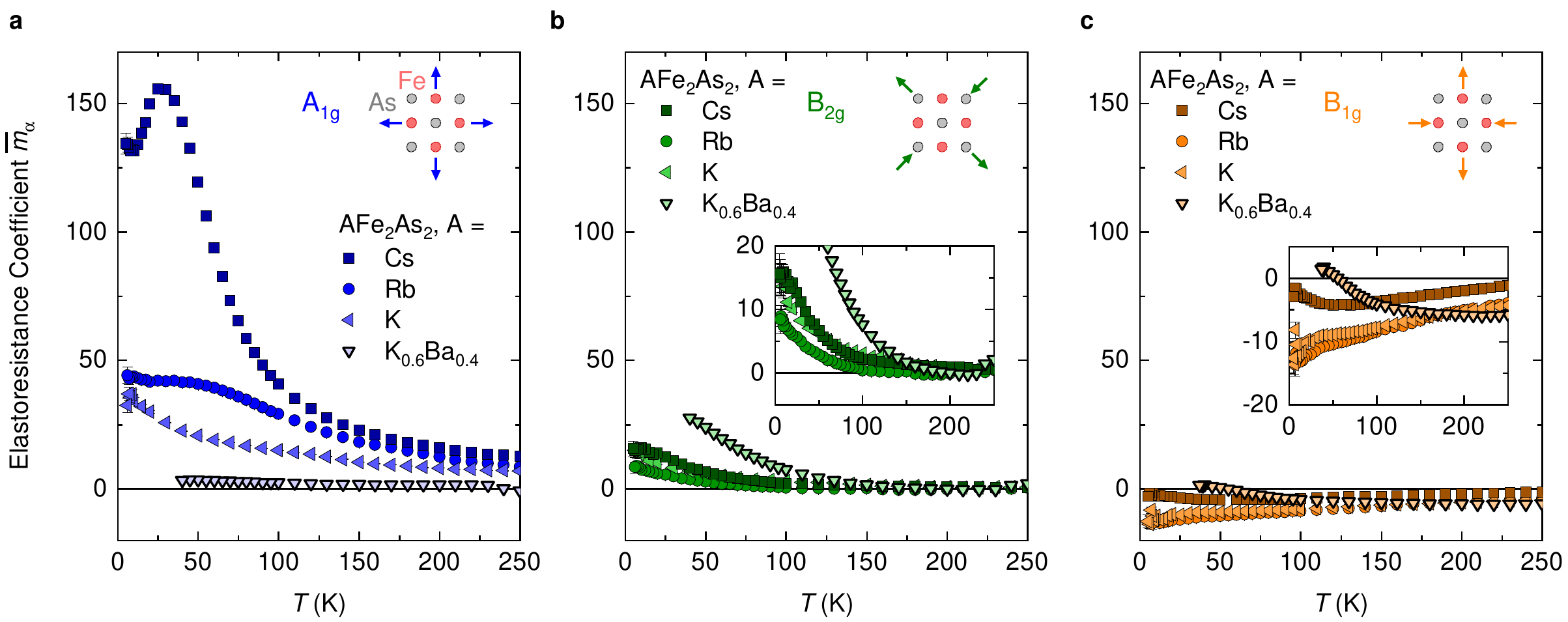}
\caption{\textbf{Symmetry-decomposed elastoresistance coefficients.}
Calculated with the redefined $\bar{m}_\alpha$ (see text).
\textbf{a}, $\bar{m}_{A_{1g}}$ calculated from $\epsilon_{xx}\|[110]$ data. \textbf{b}, $\bar{m}_{B_{2g}}$. \textbf{c}, $\bar{m}_{B_{1g}}$.
The strain symmetry channels are illustrated in the upper right of each panel.
The insets in panels \textbf{b} and \textbf{c} show a zoomed view.
The error bars take into account the uncertainty in $R_0$.
Due to large uncertainties in $R_0$ for the high-$T_c$ sample Ba$_{0.4}$K$_{0.6}$, we have plotted $m_\alpha$ instead of $\bar{m}_\alpha$ for this sample
(points with thick edges). 
}
\label{fig:irrep}
\end{figure*}

\begin{figure}[t]
\centering
\includegraphics[width=0.5\columnwidth]{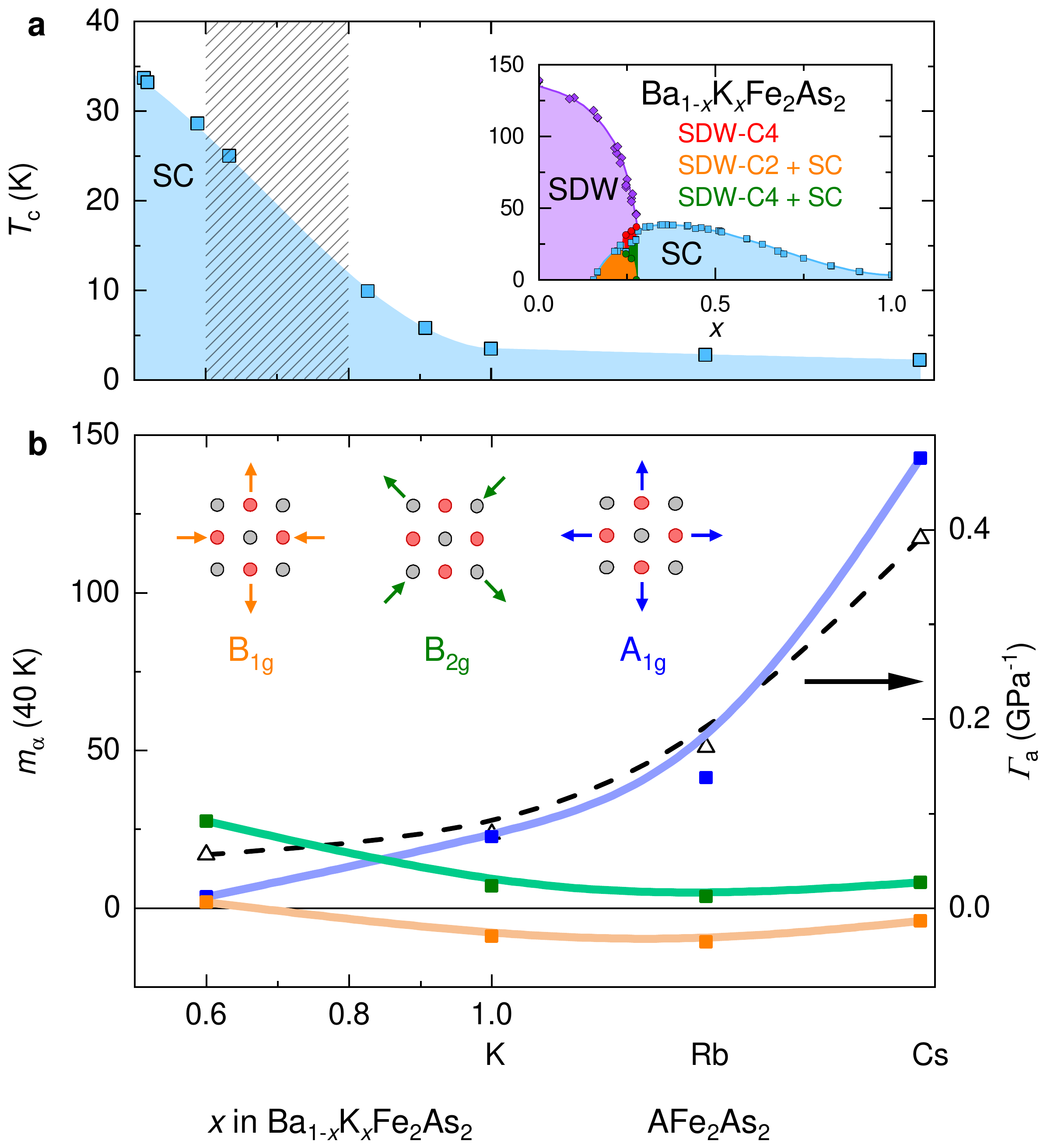}
\caption{\textbf{Evolution of the elastoresistance coefficients with chemical substitution.} 
\textbf{a} Phase diagram of the hole-doped series Ba$_{1-x}$K$_{x}$Fe$_2$As$_2$ for $x>0.3$ and 
the isoelectronic series $3d^{5.5}$ series AFe$_2$As$_2$ (A = K, Rb, Cs) on the same horizontal axis.
The hashed area denotes the doping range of a Lifshitz transition\cite{Xu_2013,Malaeb_2012,Liu_2014,Hodovanets_2014} 
and proposed broken time-reversal symmetry in the SC state\cite{Maiti_2013,Boeker_2017,Grinenko_2020}.
XY-type nematic fluctuations were suggested in the same doping range in the related system Ba$_{1-x}$Rb$_{x}$Fe$_2$As$_2$\cite{Ishida6424}.
Inset: full thermodynamic phase diagram of Ba$_{1-x}$K$_{x}$Fe$_2$As$_2$.
\textbf{b} Value of the elastoresistance coefficients $\bar{m}_\alpha$ at 40 K as a function of doping, 
for the three symmetry channels. 
The black triangles (right axis) represent the in-plane electronic Gr\"{u}neisen parameter $\Gamma$, for 
comparison with $\bar{m}_{A_{1g}}$.  Note the experimental definition\cite{Meingast_2012,Hardy_2016} $\Gamma_a\equiv\alpha_a^{\rm{elec}}/C^{\rm{elec}}=-(d\gamma/dp_a)/\gamma$ where $p_a$ is the in-plane uniaxial pressure. 
}
\label{fig:phase}
\end{figure}

We discuss first the in-plane symmetric coefficient $\bar{m}_{A_{1g}}$ (Fig. \ref{fig:irrep}\textbf{a}). 
Since the resistance reflects the electronic entropy and correlations in these materials\cite{Wu_2016,PhysRevLett.125.187001},
the increase of $\bar{m}_{A_{1g}}$ from KFe$_2$As$_2$ to CsFe$_2$As$_2$ indicates that the strain sensitivity of the electronic correlations increases as a result of this chemical substitution.
This is consistent
with the increase in the strain derivative of the 
Sommerfeld coefficient $\partial\gamma/\partial\epsilon$
observed in thermodynamic measurements\cite{Eilers_2016,Hardy_2016}.
Therefore, the origin of the large $A_{1g}$ elastoresistance is the high strain sensitivity of the electron correlations associated with the orbital-selective Mott behavior. As a consequence, the temperature and substitution dependence of the $A_{1g}$ coefficient show a strong similarity to that of the thermal expansion 
coefficient\cite{PhysRevLett.125.187001,Hardy_2016} (Supplementary Information).

We turn now to the coefficients $\bar{m}_{B_{2g}}$ (Fig. \ref{fig:irrep}\textbf{b})
and $\bar{m}_{B_{1g}}$ (Fig. \ref{fig:irrep}\textbf{c}) corresponding to symmetry-breaking strain. They are proportional to the nematic susceptibility in the respective symmetry channels\cite{Kuo958}, though the proportionality constant may become temperature and material dependent \cite{boehmer2020evolution}. Note that the underdoped BaFe$_2$As$_2$ compounds are characterized by a diverging  $\bar{m}_{B_{2g}}$. 
For all samples, $\bar{m}_{B_{2g}}$ is near zero at high temperature and displays divergent increase on cooling. 
In AFe$_2$As$_2$ (A = K, Rb, Cs), $\bar{m}_{B_{2g}}$ is smaller than in Ba$_{0.4}$K$_{0.6}$Fe$_2$As$_2$ but has a similar temperature dependence, 
strongly suggesting that we are observing the remnants of the $B_{2g}$ nematic fluctuations found in lightly-doped BaFe$_2$As$_2$. 
The temperature dependence of the coefficient $\bar{m}_{B_{1g}}$ does not show a divergence with
decreasing temperature, an indication that the AFe$_2$As$_2$ (A = K, Rb, Cs) compounds are not near a $B_{1g}$ nematic instability. 
The evolution of $\bar{m}_{B_{1g}}$ and $\bar{m}_{B_{2g}}$ show no indication of a change from $B_{2g}$ to $B_{1g}$ nematic fluctuations in hole-doped iron based superconductors, in contrast to 
previous studies\cite{Ishida6424,Liu_2019,Onari_2019,Wang_2019,Borisov_2019}. Further, nematic fluctuations of XY, as opposed to Ising, character have been proposed in the Ba$_{1-x}$Rb$_{x}$Fe$_2$As$_2$
series in the $0.6<x<0.8$ doping range\cite{Ishida6424}.
Whereas we do find that $\bar{m}_{B_{1g}}$ and $\bar{m}_{B_{2g}}$  have similar magnitude in this doping range (in Ba$_{0.4}$K$_{0.6}$Fe$_2$As$_2$ in our case), their distinct temperature dependence is inconsistent with XY nematic fluctuations.

A phase transition to a long-range ordered $B_{1g}$ nematic ordered phase below 40 K
has been proposed in RbFe$_2$As$_2$ based on a maximum in the longitudinal elastoresistance measured along [100]\cite{Ishida6424}. 
The symmetry decomposition of the elastoresistance reveals that such a maximum occurs in the $A_{1g}$ channel, which we associate with the coherence/incoherence crossover and not a phase transition. 
Consistently, the resistance and thermal expansion\cite{Hardy_2016} data show no sign of a phase transition in RbFe$_2$As$_2$ 
(Fig. \ref{fig:intro}\textbf{a} and Supplementary Figure).

Our results are summarized in Fig. \ref{fig:phase}.
In this phase diagram, we plot the values of the elastoresistance coefficients at 40 K as a function of substitution, with 
both the hole-doped series Ba$_{1-x}$K$_{x}$Fe$_2$As$_2$ and the isoelectronic
$3d^{5.5}$ series AFe$_2$As$_2$ (A = K, Rb, Cs) on the same horizontal axis. 40 K is chosen so that
the high-$T_c$ sample can be included.
The hashed region $0.6<x<0.8$ denotes the doping range where a significant change of behavior has been observed, including a 
Lifshitz transition \cite{Xu_2013,Malaeb_2012,Liu_2014,Hodovanets_2014} and proposed broken time-reversal symmetry in the superconducting state\cite{Maiti_2013,Boeker_2017,Grinenko_2020},
as well as the proposed XY nematic state in the Rb-doped series (Ba$_{1-x}$Rb$_{x}$Fe$_2$As$_2$)\cite{Ishida6424}.
Our results suggest that this doping range coincides with the emergence of an enhanced $A_{1g}$ elastoresistance. 
Recalling that $\bar{m}_{A_{1g}}$ is expected to reflect the in-plane Gr\"{u}neisen parameter $\Gamma_a$ at low temperature,
we plot $\Gamma_a$ alongside $\bar{m}_{A_{1g}}$(40 K) in Fig. \ref{fig:phase}. The agreement confirms that the $\bar{m}_{A_{1g}}$ component of elastoresistance
is associated with the strong electronic correlations observed in thermodynamic measurements.

We stress that we observe, nevertheless, the remnant of the $B_{2g}$ nematic fluctuations of the optimally-doped regime even in fully substituted samples with $3d^{5.5}$ electronic configuration. We also observe signatures of possible weak nematic fluctuations in the $B_{1g}$ channel, but we do not observe a change from $B_{2g}$-dominant to $B_{1g}$-dominant nematic fluctuations with doping and we find no evidence for a bulk nematic state in any of our samples. 
These results revise the current understanding of nematicity in hole-doped iron based superconductors and raise some new points.
First, neither $B_{1g}$ nor $B_{2g}$ nematic fluctuations depend strongly on the alkali ion size in AFe$_2$As$_2$ (A = K, Rb, Cs). In contrast, the strong correlations related to the orbital-selective Mott behavior seen in the $A_{1g}$ channel increase dramatically with increasing alkali ion size. 
The contrasting behavior indicates that the orbital-selective Mott physics does not cause the nematic fluctuations.
Similarly, the heavy-fermion superconductor URu$_2$Si$_2$ has recently been shown to have
a large symmetric elastoresistance without nematicity\cite{Wang_2020}.
Secondly, we find that the $B_{2g}$ nematic fluctuations 
are surprisingly robust, persisting at high hole doping even beyond the Lifshitz transition where the electron pocket of the Fermi surface and the corresponding nesting properties are lost, and magnetic fluctuations become incommensurate. This may be related to why the superconducting dome stretches so far in these compounds. 

To conclude, we reflect on the different physical mechanisms responsible for the observed strain-induced resistance changes. In Ba$_{0.4}$K$_{0.6}$Fe$_2$As$_2$ with a sizeable nematic susceptibility, anisotropic strains change the measured resistance by favoring the resistance of one in-plane direction over the other, as nematic order causes a resistance anisotropy.
In CsFe$_2$As$_2$, by contrast, symmetric strains directly modulate the strength of electron correlations and effective
mass causing changes in average in-plane resistance.

\section*{Methods}
\subsection*{Sample Growth}
Single crystals of CsFe$_2$As$_2$ were grown from solution using a Cs-rich self-flux in a sealed environment\cite{Wang_2013}. Cs, Fe and As were weighted in molar ratio 8:1:11, respectively. All sample manipulations were performed in an argon glove box (O$_2$ content is $< 0.5$ ppm). Molten Cs together with a mixture of iron and arsenic powder were loaded into an alumina crucible. Typically 15-20 gr of a mixture of Cs, Fe and As were used for each crystal growth experiment. The alumina crucible with a lid was placed inside a stainless steel container and encapsulated. This was done by welding a stainless screw cap to one end of the container. This has the advantage that stainless steel does not corrode due to Cs vapour at high temperatures. The stainless steel container was placed in a tube furnace, which was evacuated at 5-10 mbar and slowly heated up to 200 $^\circ$C. The sample was kept at this temperature for 10 h and subsequently heated up to 980 $^\circ$C in 8 hours. The furnace temperature was kept constant at 980 $^\circ$C for 5 h and slowly cooled to 760 $^\circ$C in 14 days and then the furnace was canted to separate the excess flux. After cooling to room temperature, shiny plate-like crystals were easily removed from the remaining ingot.
Refined crystallographic data have been presented elsewhere\cite{PhysRevLett.125.187001}. 
KFe$_2$As$_2$ and RbFe$_2$As$_2$ single crystals were obtained under similar conditions using an alkali metal / As-rich flux\cite{Hardy_2016}.

High-quality single crystals of Ba$_{0.4}$K$_{0.6}$Fe$_2$As$_2$ were grown by a self-flux technique, using FeAs fluxes, in alumina crucibles sealed in iron cylinders using very slow cooling rates of 0.2–0.4 $^\circ$C/hour. The crystals were annealed in-situ by further slow cooling to room temperature. 

\subsection*{Elastoresistance measurements}
Elastoresistance measurements were performed in a
commercial strain cell (Razorbill Instruments CS-120).  The samples
were cut with edges along the tetragonal in-plane directions
[100] or [110], with typical dimensions of $3.0 \times 1.0 \times 0.06$ mm. The samples were fixed in the strain cell
using DevCon 5 Minute Epoxy with an effective strained
sample length of 2.0 mm. The
epoxy thickness was controlled to be 50 $\mu$m below the
sample and $\sim$30 $\mu$m above the sample. 
Cigarette paper was used to ensure electrical insulation of the sample from the titanium mounting plates.
The sample strain
was measured via a built-in capacitive displacement sensor and read out on an Andeen Hagerling 2550A capacitance bridge.
Resistance was measured using a four-point method on a
Lake Shore 372 resistance bridge.
Hans Wolbring Leitsilber 200N was used to make the electrical contacts, as contacts made with 
DuPont 4929N silver paint
were found to be mechanically unstable on strain application.

Due to the large thermal expansion mismatch between our samples and the titanium body of the strain cell,
the samples will experience tension on cooling (Fig. \ref{fig:intro}\textbf{b}). 
In the most extreme case of CsFe$_2$As$_2$, the sample is expected to experience $\sim0.7\%$ tension at base temperature. 
In contrast, the parent compound BaFe$_2$As$_2$ has a thermal expansion nearly identical to titanium.
This is also true for lightly-doped compounds\cite{PhysRevB.98.245133}.
Elastoresistance has typically been measured by gluing the sample directly onto the side of a piezoelectric stack, which has a small thermal expansion, similar to titanium\cite{bohmermeingast}. 
The success of previous elastoresistance measurements on lightly doped BaFe$_2$As$_2$ samples thus depended on the fortunately similar thermal expansion of the piezoelectric stack and the sample.
However, this method is unreliable for the $3d^{5.5}$ superconductors\cite{PhysRevLett.125.187001}.

To keep the sample in an unstrained state on cooling, we progressively reduce the distance between the sample mounting plates by an amount calculated based on the known thermal expansion difference between the sample and titanium, and the measurement of a titanium calibration sample\cite{PhysRevLett.125.187001}.

To express the elastoresistance in terms of the irreducible representations $\alpha$ of the $D_{4h}$ point group,
we note that the symmetric strains $\epsilon_\alpha$ are given in terms of $\epsilon_{xx}$ by $\varepsilon_{A_{1g}}=\frac{1}{2}(\varepsilon_{[100]}+\varepsilon_{[010]})=\frac{1}{2}(\varepsilon_{[110]}+\varepsilon_{[\bar{1}10]})$, $\varepsilon_{B_{1g}}=\frac{1}{2}(\varepsilon_{[100]}-\varepsilon_{[010]})$ and $\varepsilon_{B_{2g}}=\frac{1}{2}(\varepsilon_{[110]}-\varepsilon_{[\bar{1}10]})$, where we use
the simplified notation $\varepsilon_x\equiv\varepsilon_{xx}$.
Similar expressions apply relating $R_\alpha$ with $R_{xx}$.
The symmetry-decomposed elastoresistance coefficients are then defined as $m_\alpha=(1/R_\alpha)dR_\alpha/d\varepsilon_\alpha$. 
In terms of the experimental data, these coefficients are given by
\begin{subequations}
\begin{align}
m_{A_{1g}}&=\frac{1}{1-\nu_{[100]}}\left[\frac{1}{R_{[100]}}\frac{dR_{[100]}}{d\varepsilon_{[100]}}+
\frac{1}{R_{[010]}}\frac{d R_{[010]}}{d\varepsilon_{[100]}}\right]\nonumber\\
&=\frac{1}{1-\nu_{[110]}}\left[\frac{1}{R_{[110]}}\frac{dR_{[110]}}{d\varepsilon_{[110]}}
+\frac{1}{R_{[\bar{1}10]}}\frac{dR_{[\bar{1}10]}}{d\varepsilon_{[110]}}\right]\nonumber\\
m_{B_{1g}}&=\frac{1}{1+\nu_{[100]}}\left[\frac{1}{R_{[100]}}\frac{dR_{[100]}}{d\varepsilon_{[100]}}-
\frac{1}{R_{[010]}}\frac{d R_{[010]}}{d\varepsilon_{[100]}}\right]\nonumber\\
m_{B_{2g}}&=\frac{1}{1+\nu_{[110]}}\left[\frac{1}{R_{[110]}}\frac{dR_{[110]}}{d\varepsilon_{[110]}}
-\frac{1}{R_{[\bar{1}10]}}\frac{dR_{[\bar{1}10]}}{d\varepsilon_{[110]}}\right]\nonumber.
\end{align}
\end{subequations}
The quantity $\nu$ is the sample's Poisson ratio, which relates the strain along different directions, $\nu_{[100]}=-\varepsilon_{[010]}/\varepsilon_{[100]}$, and $\nu_{[110]}=-\varepsilon_{[\bar{1}10]}/\varepsilon_{[110]}$. In general, $\nu$ is sample and temperature dependent and anisotropic ($\nu_{[100]}\neq\nu_{[110]}$).
The elastoresistance coefficients in the notation of irreducible representations
 are given in terms of the elastoresistance coefficients in the usual Voigt notation as
\begin{subequations}
\begin{align}
m_{B_{1g}}&=m_{11}-m_{12}\nonumber\\
m_{B_{2g}}&=2m_{66}\nonumber\\
m_{A_{1g}}&=(m_{11}+m_{12})-\frac{2\nu'}{1-\nu}m_{13}\nonumber\\
&=m_{A_{1g,1}}-\frac{2\nu'}{1-\nu}m_{A_{1g,2}}\nonumber
\end{align}
\end{subequations}
In the $A_{1g}$ channel, $m_{13}=m_{A_{1g,2}}$ contributes and cannot be disentangled from $m_{11}+m_{12}=m_{A_{1g,1}}$. 
Our measured elastoresistance coefficient $m_{A_{1g}}$ therefore includes the effect of $c$-axis compression (tension) which accompanies the symmetric in-plane biaxial tension (compression) because of the Poisson effect.

\begin{table}[t]
\centering
\begin{tabular}{l*{8}{c}r}
& $c_{11}$ & $c_{12}$ & $c_{13}$ & $c_{33}$ & $c_{66}$  & $\nu_{[100]}$ & $\nu'_{[100]}$ & $\nu_{[110]}$ & $\nu'_{[110]}$\\
\hline
Ba$_{0.5}$K$_{0.5}$Fe$_2$As$_2$ & 123.1 & 51.3 & 66.2 & 92.2 & 51.0 & 0.049 & 0.682 & -0.126 & 0.808  \\
KFe$_2$As$_2$            & 79.7 & 46.6 & 38.4 & 45.1 & 31.5 & 0.296 & 0.600 & -0.017 & 0.866  \\
RbFe$_2$As$_2$           & 78.3 & 40.2 & 28.4 & 53.8 &  33.2 & 0.398 &  0.318 & 0.143 & 0.453  \\
CsFe$_2$As$_2$           & 84.4 & 45.9 & 39.5 & 61.2 &  31.6 & 0.346 &  0.422 & 0.113 & 0.572 \\
\end{tabular}
\caption{Table of elastic constants (in units of GPa) and Poisson ratios (unitless), obtained from ab-initio calculations.}
\label{tab}
\end{table}

\subsection*{Calculation of Poisson ratios}
The sample Poisson ratios $\nu_x=-\epsilon_{yy}/\epsilon_{xx}$ and $\nu'_x=-\epsilon_{zz}/\epsilon_{xx}$ in terms of the 
elastic constants $c_{ij}$ are given by
\begin{subequations}
\begin{align}
\nu_{[100]}&=\frac{c_{13}^2-c_{12}c_{33}}{c_{13}^2-c_{11}c_{33}}\nonumber\\
\nu'_{[100]}&=\frac{(c_{12}-c_{11})c_{13}}{c_{13}^2-c_{11}c_{33}}\nonumber\\
\nu_{[110]}&=\frac{c_{33} (c_{11}+c_{12}-2c_{66})-2c_{13}^2}{c_{33} (c_{11}+c_{12}+2c_{66})-2c_{13}^2}\nonumber\\
\nu'_{[110]}&=\frac{4c_{13}c_{66}}{c_{33} (c_{11}+c_{12}+2c_{66})-2c_{13}^2}.\nonumber
\end{align}
\end{subequations}

To obtain rough estimates of the elastic constants, we performed ab-initio calculations in the framework of the generalized gradient approximation using a mixed-basis pseudopotential method\cite{Eilers_2016}. The phonon dispersions and corresponding inter-atomic force constants were calculated via density functional perturbation theory and the elastic constants were then extracted via the method of long waves.
For Ba$_{0.4}$K$_{0.6}$Fe$_2$As$_2$, we used values calculated for Ba$_{0.5}$K$_{0.5}$Fe$_2$As$_2$ in a virtual-crystal
approximation. 
The numerical values of elastic constants and Poisson ratios calculated therefrom are given in Table \ref{tab}.
For the purposes of calculating the elastoresistance coefficients, the Poisson ratios were taken to be temperature independent. 

\bibliography{WieckiBibTeX}

\section*{Acknowledgements}

We thank I. R. Fisher and K. Grube for valuable discussions. We acknowledge support by the Helmholtz Association
under Contract No. VH-NG-1242.

\section*{Author contributions}

P.W and A.E.B initiated the project. A.A.H. and T.W. grew single crystal samples. P.W., A.E.B and M.F. performed the experiments.  R.H. performed DFT calculations of elastic constants. P.W. analysed the data and P.W., C.M and A.E.B developed the interpretations. A.E.B. guided the project. P.W. and A.E.B wrote the manuscript with input from all authors.

\end{document}